# Real-Time Well Log Prediction from Drilling Data Using Deep Learning

Ryan Kanfar[1,2], Obai Shaikh[1,3], Mehrdad Yousefzadeh[1], Tapan Mukerji[1,3]

[1] Energy Resources Engineering Department, Stanford University, Stanford, CA 94305, USA
[2] Saudi Aramco
[3] Geophysics Department, Stanford University, Stanford, CA 94305, USA

## Abstract

The objective is to study the feasibility of predicting subsurface rock properties in wells from real-time drilling data. Geophysical logs, namely, density, porosity and sonic logs are of paramount importance for subsurface resource estimation and exploitation. These wireline petro-physical measurements are selectively deployed as they are expensive to acquire; meanwhile, drilling information is recorded in every drilled well. Hence a predictive tool for wireline log prediction from drilling data can help management make decisions about data acquisition, especially for delineation and production wells. This problem is non-linear with strong ineractions between drilling parameters; hence the potential for deep learning to address this problem is explored. We present a workflow for data augmentation and feature engineering using Distance-based Global Sensitivity Analysis. We propose an Inception-based Convolutional Neural Network combined with a Temporal Convolutional Network as the deep learning model. The model is designed to learn both low and high frequency content of the data. 12 wells from the Equinor dataset for the Volve field in the North Sea are used for learning. The model predictions not only capture trends but are also physically consistent across density, porosity, and sonic logs. On the test data, the mean square error reaches a low value of 0.04 but the correlation coefficient plateaus around 0.6. The model is able however to differentiate between different types of rocks such as cemented sandstone, unconsolidated sands, and shale.

## Introduction

Drilling parameters are available in every drilled well while wireline logs are deployed selectively and are not available at all depths. In exploratory wells, more data is acquired at the well location compared to delineation or production wells. We would like to predict rock properties from drilling parameters while drilling to help management make real time decisions about borehole wireline measurements, especially for delineation and production wells.

Multiple efforts are made in the deep learning community to predict missing well log sections from other wireline logs using time-series optimized architectures such as recurrent neural networks or more basic fully connected layers. (e.g., Zhang et al, 2018; Baneshi, 2018; Parapuram et al 2015). Identification of subsurface lithology from drilling data using a fully connected layer has also been investigated (Muazzeni and Haffar, 2009). However, no reported work map drilling parameters to geophysical logs, namely, density, porosity and sonic logs.

Drilling parameters such as rate of penetration, weight on bit, bit size, revolutions per minute, torque, flow rate, and mechanical specific energy have low linear correlation with wireline data such as density, porosity and sonic ranging from 0.2 – 0.55 as shown in figure 1 below. However, as it will be shown in the section about feature engineering, wireling logs are statistically sensitive to drilling parameters. This suggests that there is a non-linear relationship between drilling parameters and density, porosity, and sonic. Drilling parameters are also expected to be highly interactive with each other. In other words, changes in some drilling parameters such as pump pressure surely affects other drilling parameters such as flow rate and rate of penetration, etc. Since the relationship between drilling parameters and wireline logs is non-linear and interactive, we explore its deep learning potential.

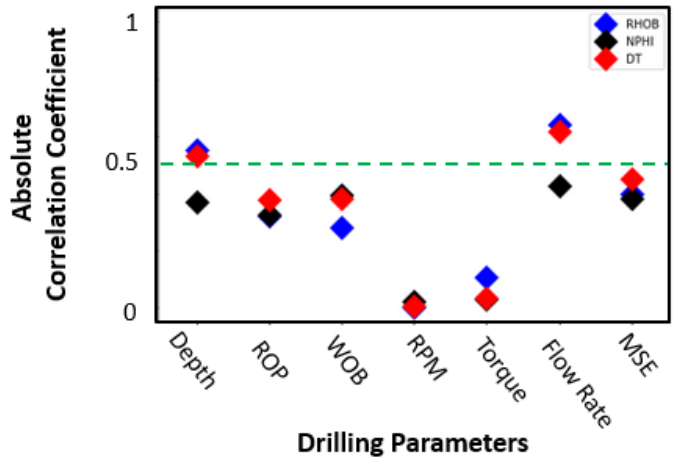

**Figure 1: Correlation coefficient of drilling paraters with density, neutron-porosity and sonic from the Volve dataset. The green dashed line is set to a correlation coefficient of 0.5. The data suggest that there is a low linear correlation between the drilling parameters and geophysical logs.**

## Data Overview

The dataset is provided by a Norwegian multinational energy company called Equinor. The data is from the Volve Field in the North Sea, which was made available since June 2018. It includes well logs, production data, seismic data, vertical seismic profile, reports, and well and seismic interpretations. The field has 24 logged wells. Unfortunately only 12 wells include density, sonic, and neutron porosity, with a total of about 89549 data points. We only used the drilling parameters to predict the geophysical logs. The drilling data includes rate of peneratration, revolutions per minute, weight on bit, torque, depth, pump pressure, flow rate, and mechanical specific energy.

The primary target in this field was the Heimdal Formation. It is a massive sandstone cemented with silica and clay. Unfortunately it was not charged. The secondary target which proved successful is an oil bearing fractured sandstones in a fluvial depositional system with a mouth bar setting called the Hugen Formation, which is Jurassic in age. The field is generally highly faulted and includes formations with diverse lithology, which makes the statistical learning process harder given the limited data. Other formations include argillaceous sandstones cemented with dolomite and kaolinite, shale, claystone with limestone stringers and limestone with grey claystone. To familiarlize ourselves with the data, a crossplot of the formations in the P-Modulus vs. neutron porosity domain is plotted in figure 2.

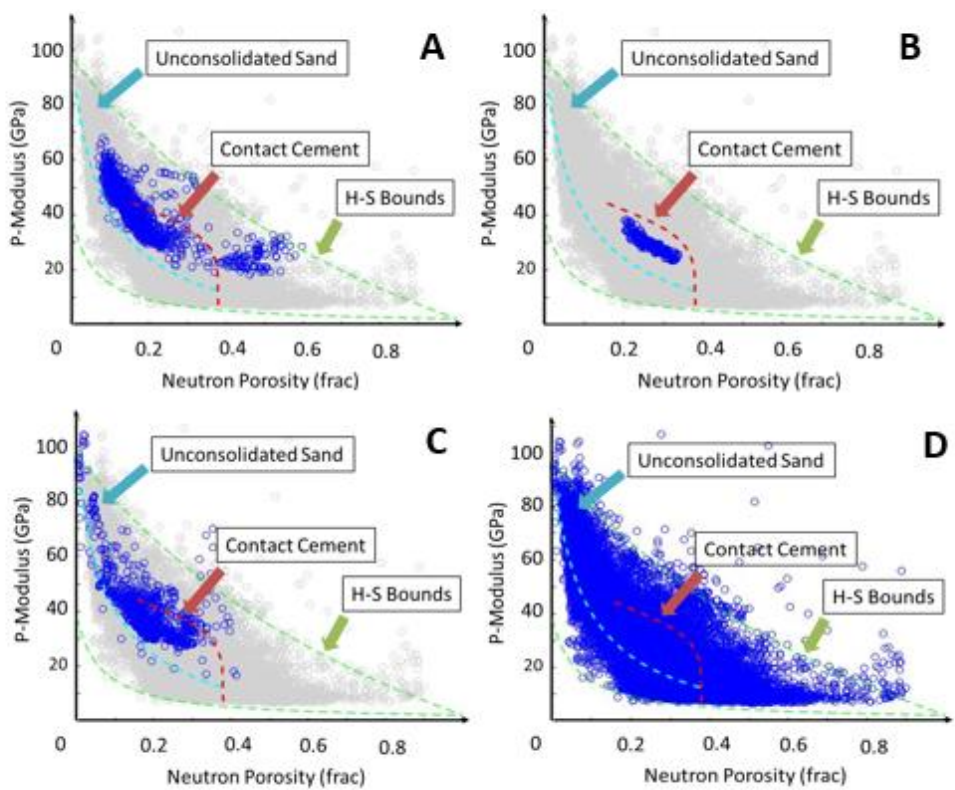

Figure 2: P-Modulus vs. neutron porosity of the 12 wells in the Volve field along with the unconsdoliated sand model, contact cement model (Dvorkin et al., 1991 and 1994) and the Hashin Shtrikman bounds (Hashin and Shtrikman, 1963.) A) Heimdal Formation: massive sandstone cemented with silica and clay. B) Hugen Formation: Oil bearing fluvial channel clean fractured sandstones in a mouth bar setting. C) Skagerrak Formation: Argillaceous sandstone in a fluvial depositional setting which is heavily cemented with dolomite and kaolinited. Poor reservoir quality rock. D) All available data plotted together showing the diversity of lithologies available in our dataset.

## Methodology

The workflow for wireline prediction from drilling parameters includes – (1) Feature Engineering (2) Data Processing and Augmentation, and (3) Architecture Selection

**Feature Engineering**

Problem decomposition is important for solving complex problems. Brute force approaches such as end-to-end deep learning is theoretically sound but hard to realize. For example, if we would like to classify the face of a human given an image, we first have to detect early in the network edges on the face to resolve features such as eyes, nose, and mouth, in order to finally recognize the face. Noteably, it would be easier to learn the mapping function if we remove features such as the background, which is not relevant to the face. It is thus important to select features in our input that are influential to the response variable of interest. The goal of feature selection is to pick influential features to our predicted variables and remove redundant or irrelevant features in order to reduce the dimensionality of the problem. For example, highly correlated features are redundant because they exhibit the same underlying information and can be removed without any loss of information.

Firstly, some drilling parameters are discrete and hard to interpret such as bit size. One way to incorporate bit size in a meaningful way is to use Mechanical Specific Energy as a collective feature instead of bit size. Mechanical Specific Energy is calculated from drilling parameters and it is the energy required to drill a unit volume of rock. MSE is calculated as follows:

$$MSE = \frac{480 \cdot Torque \cdot RPM}{d_{bit}^2 \, ROP} + \frac{4 \, WOB}{\pi \, d_{bit}^2}$$

We use Distance-based Global Sensitivity Analysis (Park et al., 2016) in order to study the sensitivity of wireline logs to drilling data and select the optimum features for predicting each of the density, porosity, and sonic logs. Figure 3 shows the sensitivity of porosity to drilling parameters and other wireline logs such as density, gamma ray and sonic. Figure 4 shows the interactive sensitivity of drilling parameters when porosity is the response variable. As expected, porosity is most sensitive to sonic, density and gamma ray. Interestingly, it is also sensitive to drilling parameters, although less sensitive than wireline logs. The sensitivities of porosity to drilling parameters are almost the same for all parameters except revolutions per minute, which is found to be the least sensitive parameter. Although we statistically proved our response variable, porosity, to be sensitive to our predictor variables, the drilling parameters, we cannot with confidence select the best predictor parameters for porosity using this sensitivity tornado plot.

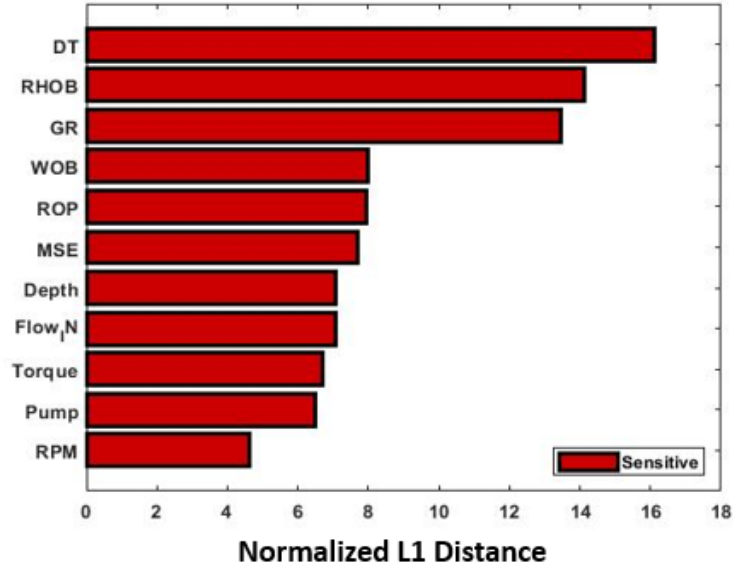

**Figure 3: Distance-based Generalized Sensitivity Analysis sensitivity plot of neutron porosity as a response variable. The x-axis is the L1 normalized sensitivity measure. Under a significance level of 0.05, any distance above 1 is considered sensitive.**

Drilling parameters are sometimes controlled and only make sense when viewed together with other drilling parameters. For example, the drilling design might plan for the rate of penetration to be a particular value over a certain period of time, while other parameters are changed to maintain the rate of penetration planned by the drilling design. We may also want to introduce either flow rate or pump pressure but not both as a feature in our inputs because the pump pressure directly affects the flow rate. In other words, interactions between parameters may indicate redundancy of information. Thus, studying the interaction between features and their effect on the response variables is important for picking influential features. The interactive sensitivity plot using DGSA helps us make decisions about this selection.

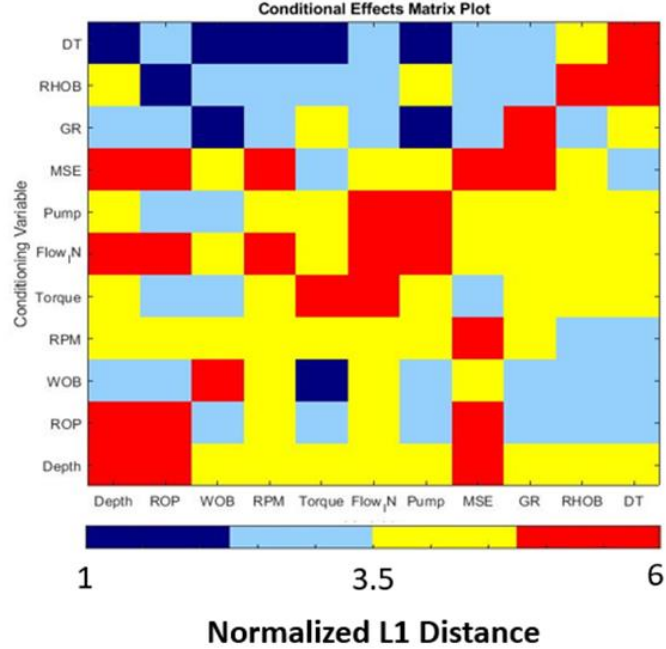

**Figure 4:** Distance-based Generalized Sensitivity Analysis interactive sensitivity for neutron porosity as a response variable. The color is the L1 normalized sensitivity measure. Under a significance level of 0.05, any value above 1 is considered interactively sensitive. The max value is capped at 6 for viewing purposes.

The diagonal of the conditional interactive sensitivity matrix plot is basically the sensitivity plot of figure 1. In general, we observe that the drilling parameters are highly interactive with each other but not when conditioning to wireline logs. As expected, however, all parameters are conditionally sensitive to depth. Additionally, flow rate and pump pressure are highly interactive. From the sensitivity and conditional interactive sensitivity matrix plot and empirical experiments we decided to use depth, rate of penetration, weight on bit, flow rate, and mechanical specific energy as input features in our model for predicting porosity and density while we use rate of penetration, weight on bit, torque, and mechanical specific energy for compressional sonic.

**Data Processing and Augmentation**

Data processing includs removal of missing data and standardization by subtracting the mean and dividing by the standard deviation. The standardization is done in order to optimize our learning through gradient descent.

In order to compensate for the diverse geology and lack of representative data, we use a striding window to augment our data. In other words, we set a window size, preferably a window size that is geologically meaningful, and stride by a fixed depth interval (for example 25 ft window and 5 ft stride respectively.) While before augmentation our data has 89549 points, after augmentation it is 834000 points, an increase by an order of magnitude.

The test data has to be carefully selected not to completely overlap the training data. Depending on the network, using this augmentation method, we could treat every window as an example instead of every point, which is more representative of the geology. Taking our striding window to be 5 points, where each point is about 1 feet, before augmentation we have 1790 examples and after augmentation we got 16680 examples. Our data dimension thus becomes: (16680, 50, 5), where the dimensions are (number of examples, time steps, channels). The number of channels are the different drilling logs used to predict our geophysical logs. The drilling logs we use are different for each predicted variable depending on the DGSA analysis and empirical experiments. We take 5% of the data to be our test data. This setting however will mean that there's some overlap between training and test data depending on the stride we choose to use, which is fine as long as there's no complete overlap between test and training data. In other words, whole test examples have not been trained on the network.

**Architecture Selection**

*Proposed Model: Inception-Based CNN-TCN*

Convolutional Neural Networks has been gaining a lot of momentum recently, especially for computer vision applications. Temporal Convolutional Networks have been shown to even outperform Recurrent Neural Network and Long Short-term Memory models on sequence based tasks and are recommended to be the starting point for sequence problems (Bai et al. 2018). CNNs are generally effective because their parameters are shared and their connections are sparse, which is a good model for geology. In other words, the input shares each filter striding along its values and every output depends only on the cross correlation of the parameters of the striding filter with the input. The size of these filters can be designed to pick up different low level features at the beginning of the network. CNNs can also preserve memory using dilations.

The inception model is a regular CNN that has layers that maintain a convolutional output shape that is similar to the input shape. This is done by padding the input with the mean, which in this case is zero because the data is standardized. By doing this, we are able to convolve multiple sized filters with the input and generate multiple outputs of similar size. We can then concatenate the output volumes to generate one deep volume as shown in figure 5. The inception model was developed by Google for a computer vision application (Szegedy et al. 2014.) The idea behind the inception model is making the network deeper by going wider instead in order to avoid vanishing gradients. Our hope for the Inception-based Convolutional Neural Network is to capture geologic patterns at different scales by convolving the input with different sized filters.

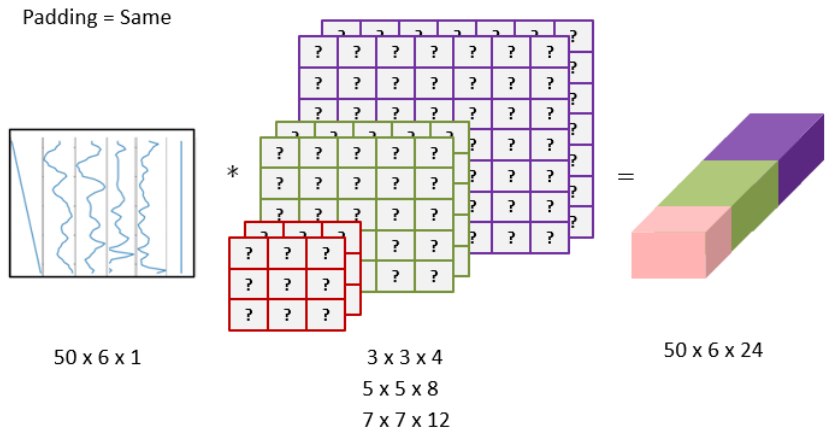

**Figure 5: Schematic showing Inception-based Convolutions. "Same" padding ensures that the output has the same dimensions as the input, which can be concatenated as shown. The third dimension is the number of filters used.**

We implement a 1D convolution instead of a 2D convolution by treating the drilling features as channels. 2D convolutions are more biased and unnecessarily computationally expensive due to the required padding across the input image in order implement the inception model. We also use a number of filters across the network that 1x1 in size to reduce the dimensionality and number of parameters by reducing the number of filters in the output. The details of the empirically optimized network for the drilling problem is shown in table 1. The inception layers are empirically found to be best placed in the middle of the network, after resolving for low level features with cascaded convolutions to the input.

We follow the Inception-Based CNN with a Temporal Convolutional Network. Upon experimentation, we found that sequence based models, such as TCN and Long Short-term Memory models tend to learn the high frequency patterns in the data but not the low frequencies. The opposite is true for regular CNNs and Inception-Based CNNs. Thus, we decided to combine the two models. We chose TCN over LSTM because it was much more stable to train.

The TCN model is a 1D convolutional network that uses residual blocks and dilated convolutions (Bai et al. 2018). Dilation simply means that we skip a number of data points in the input when we compute the cross correlation. The dilations can be set to increase exponentially with depth thereby increasing the receptive field of the activations. This approach increases the networks memory access in long sequences. TCNs are initially made to be causal with no memory leaks from the future to the past; it is designed to predict the future using only the past. In other words, every step depends on the previous steps only. Modifications were made to use the future and past to predict each time-step as shown in figure 6.

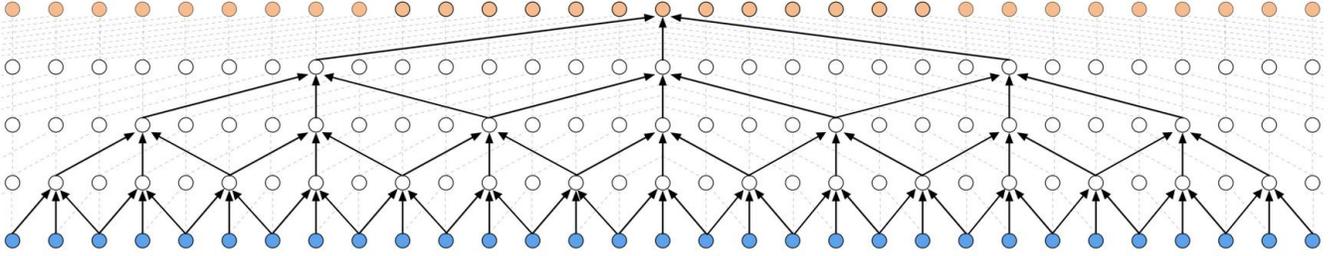

**Figure 6: Schematic showing 1 block of non-causal TCN. In this modiefied version, the output activations depend not only on the past, but also the future. The filter size above is 3 and the dilation is [1 2, 4 8].**

In the learning we make use of batch normalization, Adam optimization and drop out to generalize training predictions to test predictions. The batch size is set to 128. We also experiment with the cost function by minimizing the mean square error, minimizing the deviations of correlation coefficient and a weighted combination of the two. MSE is found to be the best cost function for our problem.

| Layer | 1 D Filter Size | Output Shape | Number of Parameters |
|---|---|---|---|
| Input | - | (50,5) | |
| Conv1D | 3 | (47,16) | 256 |
| Conv1D | 5 | (44,32) | 2592 |
| Max Pool (s = 2) | 3 | (21,32) | 0 |
| Inception1* | 3 | (21,96) | 3104 |
|  | 5 |  | 5152 |
|  | 7 |  | 7200 |
| Conv1D | 1 | (21,32) | 3104 |
| Inception2* | 3 | (21,192) | 6208 |
|  | 5 |  | 10304 |
|  | 7 |  | 14400 |
| Conv1D | 1 | (21, 64) | 12352 |
| Max Pool (s = 2) | 3 | (8, 64) | 0 |
| TCN* | 3 | (512, 3) | 426 |
| Dense | - | 256 | 393472 |
| Dense |  | 50 | 12850 |
|  |  |  | Total: 471420 |

**Table 1: Details of the architectures including layering, filter size, output shape and number of parameters. Each inception module includes three convolutions of different filter sizes. The output of the inceptions are then concatenated to generate the output shape. The TCN layer consists of one residual block with dilations set to be: [1, 2, 4, 8, 16, 32], one for each layer.**

# Results and Discussion
**Inception-based CNN**

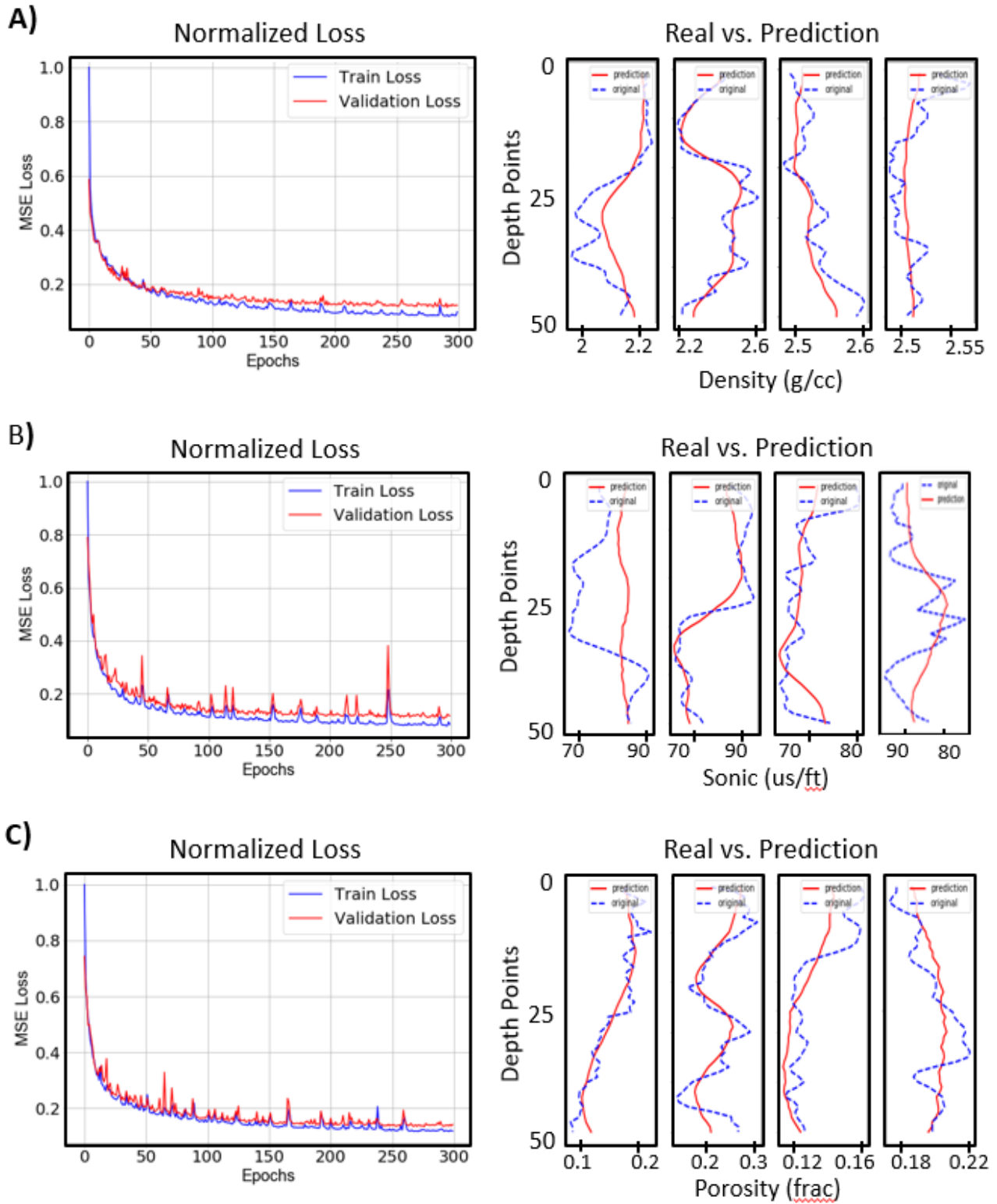

Figure 7: Performance of Inception-based CNN on test data. The loss is normalized by the maximum loss. Actual MSE values are given in table 2. The predicted variables are: A) Density, B) neutron porosity, C) compressional sonic. In the second panel, blue is the true test data and red are the predictions. Each of the 4 predictions in the second panel is a random example.

**Inception-Based CNN-TCN**

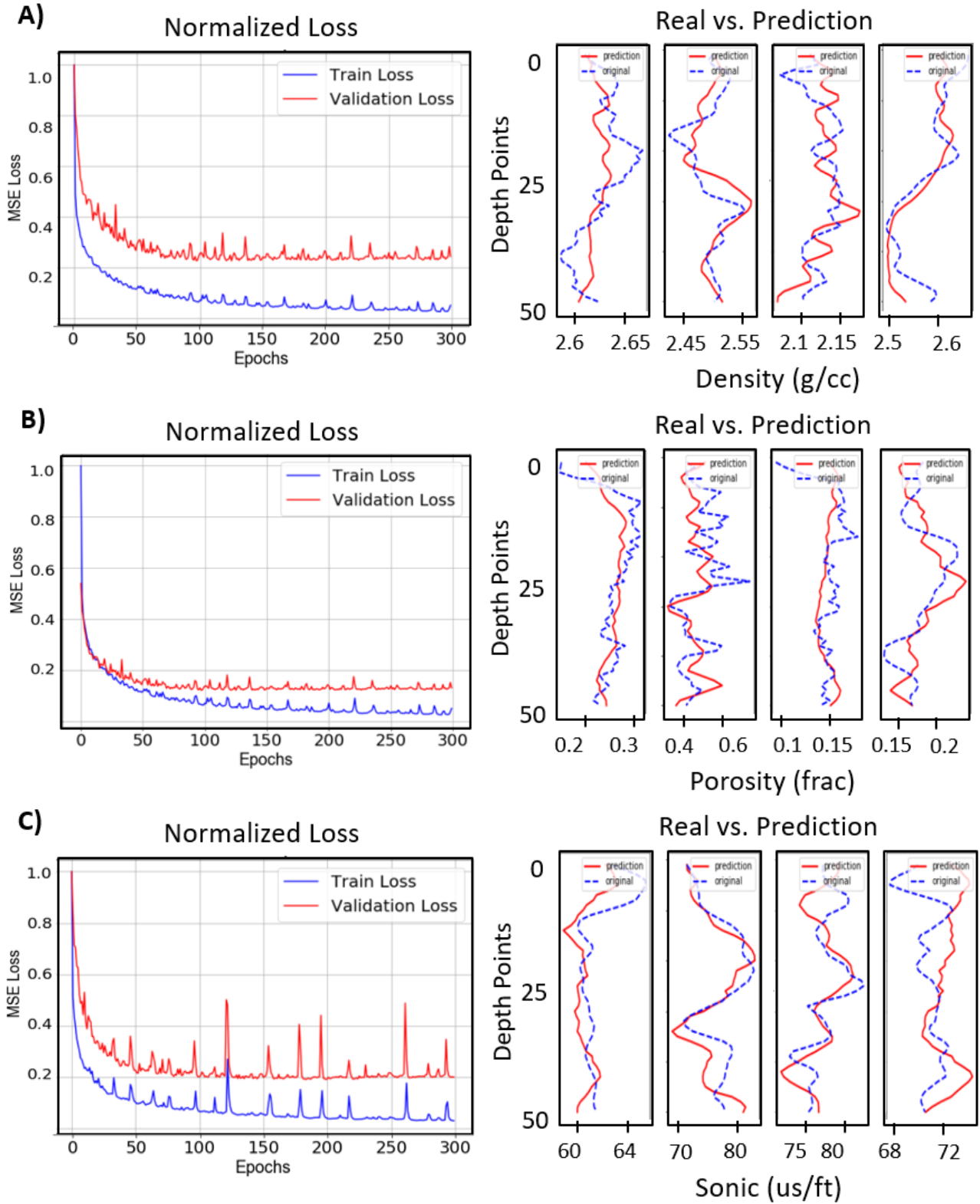

Figure 8: Performance of Inception-based CNN-TCN on test data. The loss is normalized by the maximum loss. Actual MSE values are given in table 2. The predicted variables are: A) Density, B) neutron porosity, C) compressional sonic. In the second panel, blue is the true test data and red are the predictions. Each of the 4 predictions in the second panel is a random example.

## Quality Check of Predictions

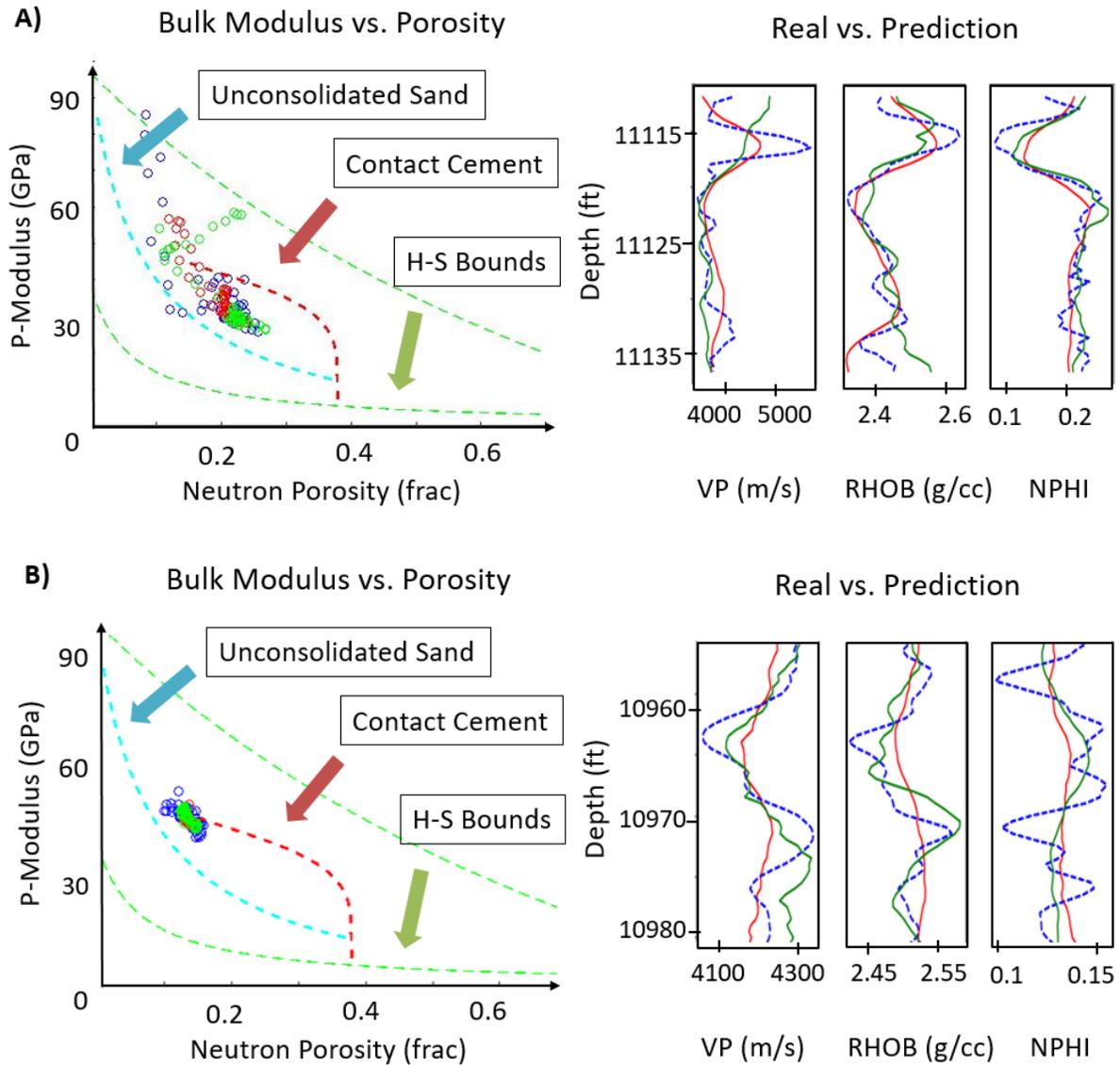

**Figure 9: Model performance. Blue, red and green points and curves are the original test data, inception-based CNN predictions, and the inception-based CNN-TCN predictions, respectively. The data in the right panel is used to generate the left panel. The crossplot has generic rock physics models such as the unconsolidated sand model, contact cement model and the Hashin-Strikman bounds as reference for understanding the data.**

| Model | Data Type | Metric | Density | NPHI | DTC |
|---|---|---|---|---|---|
| Inception-based CNN | Train | MSE | 0.08 | 0.09 | 0.11 |
| | | $\rho$ | 0.63 | 0.63 | 0.59 |
| | Test | MSE | 0.12 | 0.11 | 0.13 |
| | | $\rho$ | 0.59 | 0.6 | 0.53 |
| Inception-based CNN-TCN | Train | MSE | 0.02 | 0.02 | 0.01 |
| | | $\rho$ | 0.44 | 0.52 | 0.78 |
| | Test | MSE | 0.05 | 0.4 | 0.04 |
| | | $\rho$ | 0.41 | 0.47 | 0.55 |

**Table 2: Model performance on both test and training data based on two metrics: correlation coefficient and mean square error.**

We can see from the predictions in figure 6 that the inception-based CNN mostly learns the low frequencies of the data. We find that by following the network with a sequence based model, in this case a TCN block, we also learn the higher frequencies. This result was hypothesized based on the results of TCN and LSTM models where only the high frequencies were learned from the data. Initial look to the outputs looked like random noise, but we decided to test our hypothesis by combining the two networks. One explanation for sequence models mainly learning high frequencies in our problem is that the optimized weights in the LSTM and TCN models are shared across the whole network. In other words, because of the memory property, the network tries to optimize predictions that have high receptive fields, which could become problematic if the predictions share some qualities, in this case a high frequency behavior.

Although the correlation coefficients of predictions on test data are low, ranging between 0.4 and 0.6 depending on the model, trends seem to be captured in both the inception-based CNN and inception-based CNN-TCN models as seen in figure 7 and 8. Furthermore, the predicted trends are physically consistent across the predicted variables in both networks, as shown in figure 9. For example, an increase in density is usually coupled with an increase in compressional sonic and a decrease in porosity. Finally, figures 9A and 9B show a crossplot of P-modulus vs. porosity along with rock physics models and bounds. Our predictions lie within their expected lithological range. In other words, sandstones are not predicted as shale or limestone and vice versa. Figure 9A and 9B are believed to be the clean fractured sandstone of the Hugen formation and the massive sandstone cemented with silica and clay of the Heimdal formation. Our predictions could also be given the same interpretation.

Minimizing the mean square error cost function results in predictions better than maximizing the correlation coefficient or minimizing both the MSE and maximizing the correlation coefficient of the predictions with different weights. The normalized loss shows a +90% decrease in loss over 300 epochs. Even with regularization, following TCN to CNN makes it harder for the network to generalize to the test data as shown in the loss deviations between training and validation data in figure 7. Correlation coefficients of 0.7 on the training data suggest that there is more room for improvement in the architecture. More experimentation with the network could yield better results.

## Conclusion

We introduced a workflow for predicting petrophysical logs from drilling parameters. The workflow includes data augmentation and feature engineering using DGSA. We also explain the reasons behind our model design, which outperforms other conventional networks. The results show that given the right model, it is feasible to predict geophysical logs from drilling data. Trends are captured and are physically consistent across density, porosity and compressional sonic data. The predictions are also within reasonable lithological ranges. Given more data and model complexity, we expect to predict completely blind wells.

## Acknowledgements

We would like to thank the Stanford Rock Physics & Borehole Geophysics Project, Stanford Center for Earth Resources Forecasting, and Saudi Aramco for their support and valuable discussions. We also extend gratitude to Equinor for making their data available.